\documentstyle [fleqn,12pt,epsf] {article}
\def\double{\baselineskip 24pt \lineskip 10pt}

\begin{document}
\centerline{\large\bf Doppler Peaks in the Angular Power Spectrum}
\centerline{\large\bf of the
   Cosmic Microwave Background:}
\centerline{\large\bf A Fingerprint of Topological Defects.}
\vskip24pt
\baselineskip=10pt
\centerline{Ruth Durrer$^{\star}$, Alejandro Gangui$^{\dagger\ddagger}$ and
	Mairi Sakellariadou$^{\star}$}
\vskip10pt
\centerline{\it $^{\star}$ Universit\'e de Gen\`eve}
\centerline {D\'epartement de Physique Th\'eorique}
\centerline{\it 4, quai E. Ansermet}
\centerline{\it CH-1211 Gen\`eve 4, Switzerland}
\vskip 11pt
\centerline{\it $^{\dagger}$ SISSA -- International School for
                                      Advanced Studies}
\centerline{\it Strada Costiera 11, 34014 Trieste, Italy}
\vskip 11pt
\centerline{\it $^{\ddagger}$ ICTP -- International Center for
                                      Theoretical Physics}
\centerline{\it P.~O.~Box 586, 34100 Trieste, Italy}
\vskip 44pt
\begin{abstract}
\vskip 10pt
The Doppler peaks (Sacharov peaks) in the angular power spectrum of the
cosmic microwave background anisotropies,
are mainly due to coherent oscillations in
the baryon radiation plasma before recombination.
Here we present a calculation of the Doppler peaks for perturbations
induced by global textures and cold dark matter.
We find  that  the height of the first Doppler peak is smaller than
in standard cold dark matter models,
and that its position is shifted to $\ell\sim 350$.
We believe that our analysis can be easily extended to
other types of global topological defects and general global scalar
fields.

\vspace{1cm}
\end{abstract}
\vspace{1cm}

\double

Presently there are two main classes of models to explain  the origin
of large scale structure formation. Initial perturbations can either
be due to quantum fluctuations of a scalar field during an inflationary
era\cite{Stein}, or they may be seeded by topological defects formed
during a symmetry breaking phase
transition in the early universe\cite{Kibble}.
The CMB anisotropies are a powerful tool to discriminate among
these models  by  purely linear analysis.
Usually CMB anisotropies are parameterized in terms of
$C_\ell$'s, defined as the coefficients in the expansion of the angular
correlation function
\[ \langle{\delta T\over T}({\bf n}){\delta T\over T}({\bf n}')
\rangle\left|_{{~}_{\!\!({\bf n\cdot n}'=\cos\vartheta)}}\right. =
  {1\over 4\pi}\sum_\ell(2\ell+1)C_\ell P_\ell(\cos\vartheta). \]
For scale invariant spectra of perturbations $\ell (\ell + 1) C_\ell$
is constant on large angular scales, say $\ell \stackrel{<}{\sim} 50$.
Both inflation and topological defect models lead to approximately scale
invariant spectra on large scales.

Large scale CMB anisotropies are mainly caused by inhomogeneities
in the spacetime geometry via the Sachs--Wolfe (SW) effect\cite{SW}.
On smaller angular scales
($0.1^\circ\stackrel{<}{\sim}  \theta\stackrel{<}{\sim}  2^\circ$)
the dominant contribution comes from coherent oscillations in the
baryon--radiation plasma prior to recombination.
On even smaller scales  the anisotropies
are damped due to the finite thickness of the recombination shell,
as well as by photon diffusion during recombination (Silk damping).

Disregarding Silk damping, gauge invariant linear perturbation
analysis leads to\cite{RuthReview}
\begin{equation}
{\delta T\over T} = \left[
 - {1\over 4}D_g^{(r)} -V_jn^j -\Psi+\Phi\right]_i^f +
\int_i^f (\Psi' - \Phi' ) d\tau  ~,
\label{dT} \end{equation}
where $\Phi$ and $\Psi$ are quantities describing the perturbations
in the geometry and $\bf V$ denotes the peculiar velocity of
the baryon fluid with respect to the overall Friedmann expansion.
$D_g^{(r)}$ specifies the
intrinsic
density fluctuation in the radiation fluid.
There are several gauge invariant variables which describe
density fluctuations; they all differ on super--horizon
scales but coincide inside the horizon.
Below we use another quantity,
$D_r$, for the radiation density fluctuation. The variables $D_g^{(r)}$
and $D_r$ are defined in Eq.~(II.5.28b) and Eq.~(II.5.27b) of ref. \cite{KS}
respectively, for an arbitrary matter component $\alpha$. Here $_r$ stands
for the coupled baryon radiation fluid\footnote{ Actually
$D_r = (\delta\rho_r +\delta\rho_b)/(\rho_r+\rho_b)$ is not quite the
variable for the temperature fluctuation,
$\delta T/T = (1/4)\delta\rho_r/\rho_r$. A short calculation shows that $D_r$
is about 5\% smaller than $\delta\rho_r/\rho_r$.}.
Since the coherent oscillations giving rise to the Doppler peaks
act only on sub--horizon scales, the choice of this variable
is irrelevant for our calculation.

$\Phi$ , $\Psi$ and $D_g^{(r)}$ in Eq.~(\ref{dT}) determine the
anisotropies on large angular scales\footnote{One might think that
$D_g^{(r)}$ leads {\sl just} to coherent oscillations of the baryon
radiation fluid, but this is not the case.
Note that, e.g., for adiabatic CDM models without source term
one can derive
$(1/4) D_g^{(r)}= - (5/3)\Psi$ on super--horizon scales.
Since for CDM perturbations, $\Phi=-\Psi$ and  $\Psi' \simeq 0$,
 the usual SW result  $\delta T / T =
(1/3)\Psi ({\bf x}_{rec}, t_{rec})$ is recovered.
Neglecting $D_g^{(r)}$, the result would be
$2\Psi$ and therefore wrong by a factor of 6!},
and have been calculated for both inflation and defect
models \cite{St,BR,PST,DZ}.
Generically, a scale invariant spectrum is predicted and thus the
SW calculations yield mainly a normalization for the
different models.
On the other hand the amplitude of the Doppler peak,
which most probably will be measured in the near future,
might  be an important discriminating tool between them.
In this {\sl Letter} we present a computation for the Doppler
contribution from global topological defects;
in particular we perform our analysis for $\pi_3$--defects,
textures \cite{Tu}, in a universe dominated by cold dark matter (CDM).
We believe that our main conclusion remains valid for all global defects.

The Doppler contribution to the CMB anisotropies is approximately given
by\footnote{In principle this Doppler term has to be added to the SW
contribution. But the SW contribution decays on subhorizon scales
(like $\ell^{-2}$). At horizon scales, especially the last term in
Eq.~(\ref{dT}), the integrated Sachs--Wolfe (ISW) effect, can be important.
At $\ell=200$ it contributes about 30\% to the angular power spectrum for
standard adiabatic CDM. Neglecting it, slightly shifts the first Doppler
peak to smaller angular scales, $\ell \approx 220$, as we have found by
testing our code for  the standard adiabatic CDM model. Since we will find
here, that the first Doppler peak is lower than in this model,
the ISW contribution might be higher. However, as we shall see below, the
peak is at $\ell=365$. Therefore we expect a suppression by
$(365/200)^2\approx 3$, so that
the ISW contribution to the peak is probably not much higher. We also
neglect the contribution of the neutrino fluctuations. But since even the
dark matter fluctuations yield only about 20\% of the gravitational potential,
we expect the neutrino fluctuations, which for standard models
contribute about 20\%, to be considerably smaller.}
\begin{equation}
\left[
{\delta T \over T}({\bf x,n})
\right]^{Doppler} \approx
{1\over 4}D_r({\bf x}_{rec},t_{rec}) +
{\bf V}({\bf x}_{rec},t_{rec})\cdot {\bf n},
	\label{Doppler}\end{equation}
where ${\bf x}_{rec} = {\bf x} - {\bf n}t_0$.
In the previous formula ${\bf n}$ denotes a direction in the sky
and $t$  is conformal time, with $t_0$ and $t_{rec}$ the present
and recombination times, respectively.
Due to the inaccuracies mentioned in the previous footnotes,
Eq.~(\ref{Doppler}) tends to underestimate the amplitude of the first
Doppler peak by up to 30\%. On the other hand, we neglect Silk damping
of perturbations, which leads to a slight overestimation.
 We thus are on the safe side, if we postulate that
approximation~(\ref{Doppler}) leads to an error of less than about
30\% in the amplitude of the first Doppler peak and overestimates the
value $\ell$ of its position by less than 10\%.  The nice feature of
Eq.~(\ref{Doppler}) is that we will need only one simple scalar
component of the defect stress--energy tensor to evaluate it.

To determine $D_r$ and {\bf V} at $t_{rec}$,
we consider a two--fluid system: baryons plus radiation, which prior to
recombination are tightly coupled, and CDM.
The evolution of the perturbation variables
in a flat background, $\Omega = 1$, is described by\cite{KS}
\begin{equation} \begin{array}{lll}
V'_r +{a'\over a}V_r &=& k\Psi+ k{c_s^2\over 1+w}D_r \\
V'_c +{a'\over a}V_c &=& k\Psi  \\
D'_r - 3 w {a'\over a}D_r &=& (1+w) [ 3{a'\over a}\Psi-3\Phi'
   -kV_r - {9\over2} \left({a'\over a}\right)^2 k^{-1}
   (1+{w\rho_r\over \rho})V_r  ] \\
D'_c                 &=& 3{a'\over a}\Psi-3\Phi' - k V_c
   -{9\over2} \left({a'\over a}\right)^2 k^{-1}
   (1+{w\rho_r\over \rho})V_c
{}~, \end{array}
 \label{KSeq} \end{equation}
where subscripts $_r$ and $_c$ denote the baryon--radiation plasma
and CDM, respectively;
$D,~V$ are density and velocity perturbations;
$w=p_r/\rho_r$, $c_s^2=p'_r/\rho'_r$ and $\rho = \rho_r + \rho_c$.
The only place where the seeds enter this system is through the
potentials $\Psi$ and $\Phi$. These potentials  can be split
into a part coming from standard  matter and radiation, and a part
due to the seeds, $ \Psi = \Psi_{(c,r)} +\Psi_s$ and
$ \Phi = \Phi_{(c,r)} +\Phi_s$, where $\Psi_s$ and $\Phi_s$ are
determined by the energy momentum tensor of the seeds.
In this way,  the seed source terms will arise below\cite{RuthReview}.

{}From Eqs.~(\ref{KSeq}) we derive two second order
equations for $D_r$ and $D_c$, namely
\begin{eqnarray}
D_r''+{a'\over a}[1+3c_s^2-6w+F^{-1}\rho_c]D_r'
	 -{a'\over a}\rho_cF^{-1}(1+w)D_c' &&\nonumber \\
+4\pi Ga^2[\rho_r(3w^2-8w+6c_s^2-1)-2F^{-1}w\rho_c(\rho_r+\rho_c)
	&&\nonumber\\
+\rho_c(9c_s^2-7w)+
{k^2\over 4\pi Ga^2}c_s^2]D_r -4\pi G a^2\rho_c(1+w)D_c ~~=
{}~~(1+w)S~;&&\hspace{1cm}  \label{dr}\\
D_c''+{a'\over a}[1+(1+w)F^{-1}\rho_r(1+3c_s^2)]D_c'
	 -{a'\over a}(1+3c_s^2)F^{-1}\rho_rD_r' &&\nonumber \\
	-4 \pi G a^2\rho_cD_c -
	4 \pi G a^2\rho_r(1+3c_s^2)[1-2(\rho_r+\rho_c)F^{-1}w]D_r
&=&S~,  \label{dc}
\end{eqnarray}
where $F\equiv k^2(12\pi Ga^2)^{-1}+\rho_r(1+w)+\rho_c$ and
$S$ denotes a source term,
which in general is given by $S=4\pi G a^2(\rho +3p)^{seed}$.
In our case, where the seed is described by a global scalar field $\phi$,
we have $S=8\pi G(\phi ')^2$.
{}From  numerical simulations one finds that the average of $|\phi '|^2$
over a shell of radius $k$, can be modeled during the matter dominated era
 by\cite{DZ}
\begin{equation}
 \langle|\phi'|^2\rangle (k, t)\ =\
{{1\over 2} A\eta^2\over \sqrt {t}[1+\alpha (kt)
+\beta (kt)^2]},  \label{fit}
\end{equation}
with $\eta$ denoting the symmetry breaking scale of the phase transition
leading to texture formation.
The parameters in (\ref{fit}) are
$A\sim 3.3$, $\alpha\sim -0.7/(2\pi)$ and $\beta \sim 0.7/(2\pi)^2$.
On super--horizon scales, where the source term is important,
this fit is accurate to about $10\%$.
As we argue later, analytical estimates support this finding.
On small scales the accuracy reduces to a factor of 2.
By using this fit\footnote{Our fit is
not valid in the radiation dominated era.  There, logarithmic corrections
or a different power law might have to be applied.  Since the relevant scales
enter the horizon roughly during the matter--radiation transition, this
renders the amplitude of the corresponding fluctuations somewhat uncertain.}
 in the calculation of $D_r$ and $D_c$ from
Eqs. (\ref{dr}), (\ref{dc}) we effectively neglect the
time evolution of phases of $(\phi ')^2$;
the incoherent evolution of these phases may smear out subsequent
Doppler peaks\cite{alb}, but will not affect substantially the
height of the first peak.

{}From  $D_r$ and $D_r'$ we calculate the Doppler contribution to the
$C_{\ell}$'s according to
\begin{equation}
C_{\ell} = {2\over \pi} \int dk \left[{k^2\over 16}|D_r(k,t_{rec})|^2
j_{\ell}^2(kt_0) + (1+w)^{-2} |D_r'(k,t_{rec})|^2  (j_{\ell}'(kt_0))^2\right] ,
\label{cl}
\end{equation}
where $j_\ell$ denotes the spherical Bessel function of order $\ell$ and
$j'_\ell$ stands for its derivative with respect to the argument.
The angular power spectrum
$\ell (\ell + 1) C_{\ell}$  yields the Doppler peaks.

In order to solve Eqs. (\ref{dr}), (\ref{dc}) we need to specify
initial conditions.
For a given scale $k$ we choose the initial time $t_{in}$ such that the
perturbation is super--horizon and the universe is radiation dominated.
In this limit the evolution equations reduce to
\begin{eqnarray}
D_r''-{2\over t^2} D_r &=&{4\over 3}{A\epsilon \over \sqrt {t}} ~; \\
D_c''+{3\over t} D_c'-{3\over 2t}D_r'-{3\over 2t^2}D_r  &=&{A\epsilon \over
\sqrt {t}} ~,
\end{eqnarray}
with particular solutions
\begin{equation}
D_r\ = \ -{16\over 15}\epsilon A t^{3/2}\ ;\
D_c\ = \ -{4 \over  7}\epsilon A t^{3/2}\ .
\end{equation}
In the above equations we have introduced
$\epsilon\equiv 4\pi G\eta^2$, the only free parameter in the model.
We consider perturbations  seeded by the texture field, and
therefore it is incorrect to add a homogeneous growing mode to the
above solutions.
With these initial conditions, Eqs. (\ref{dr}), (\ref{dc}) are easily
integrated numerically, leading to the spectra for $D_r(k, t_{rec})$
and $D_r '(k, t_{rec})$ [{\it see}, Fig. 1].

\begin{figure}[htb]
\centering
\epsfysize=8.5cm
\epsffile{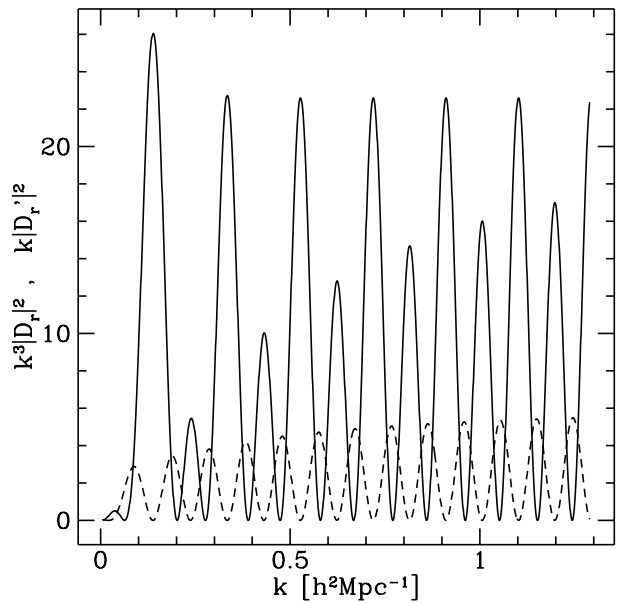}
\caption[1]{\small The dimensionless power spectra, $k^3|D_r|^2$
(solid line)
and $k|D'_r|^2$ (dashed line) in units of $(A\epsilon)^2$,
are shown as functions of $k$.
These are exactly the quantities which enter in the expression for the
$C_\ell$'s. We set $h=0.5~,~\Omega_B=0.05$ and
 $z_{rec}=1100$. }
\end{figure}

Integrating Eq. (\ref{cl}), we obtain the Doppler contribution to the
CMB anisotropies [{\it see}, Fig. 2].
For $\ell <1000$, we find three peaks located
at $\ell=365$, $\ell=720$ and $\ell = 950$. Silk damping, which we
have not taken into account here, will substantially decrease the height
of the second and even more  that of the third peak.
The integrated Sachs Wolfe effect, which also has been neglected, will
shift the position of the first peak to somewhat larger scales, lowering
$\ell_{peak}$ by (5 -- 10)\% and increasing its amplitude by less than
30\%, as we argued above.

Our second result regards the amplitude
of the first Doppler peak, for which we find
\begin{equation}
	\ell(\ell+1)C_\ell
\left|_{{~}_{\!\!  \ell\sim 365 }}
= 5\epsilon^2~.
	\right. \label{clfirstpeak}\end{equation}
As a consequence of the remark in
footnote 4, the above numerical result has to be taken with a grain of salt.
It is interesting to notice that the position of the first peak is
displaced by  $\Delta \ell\sim 150$  towards smaller
angular scales than in inflationary models \cite{St}.
This is due to the fact that our solution represents a
combination of the growing and decaying modes, and only once the perturbation
enters the horizon and the source term becomes negligible, the decaying mode
starts to decay. This is manifest in the difference in the growth of
super--horizon perturbations, which is $D_r \propto t^{3/2}$ in our case,
and $D_r \propto t^2$ for inflationary models, where on all
 scales only the growing mode is present.

\begin{figure}[t]
\centering
\epsfysize=8.5cm
\epsffile{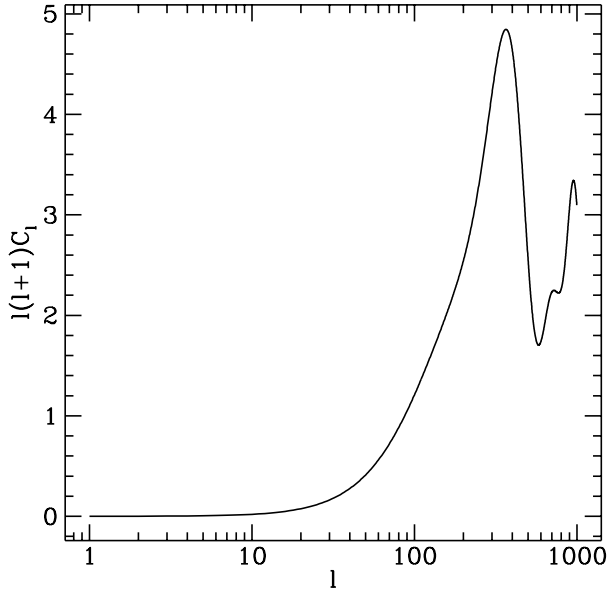}
\caption[2]{\small  The  angular power spectrum for the Doppler
contribution to
the  CMB anisotropies is shown in units of $\epsilon^2$. We set
cosmological parameters $h=0.5~,~\Omega_B=0.05$ and $z_{rec}=1100$ }
\end{figure}

One may understand the height of the first peak from the following
analytic estimate: matching the sub--horizon with the super--horizon
solutions of Eq.~(\ref{dc}), in the matter dominated era, one finds
$D_c \sim - 0.4A\epsilon(k/2\pi)^{1/2}t^2$.
 From Eq.~(\ref{dr}) we then obtain in this limit
	$D_r\approx A\epsilon k^{-3/2}$.
Plugging this latter value into Eq.~(\ref{cl}) we get roughly
$\ell(\ell+1)C_\ell \sim (A\epsilon)^2$
for the height of the first peak. This agrees, within a factor 2 with the
numerical result given in Eq.~(\ref{clfirstpeak}).

Let us now compare our value for the Doppler peak with the level of the
SW plateau. Unfortunately, the numerical value for the SW amplitude
is uncertain within a factor of about 2, which leads to a factor 4
uncertainty in the SW contribution to the power spectrum: Refs.~\cite{BR,PST}
and Ref.~\cite{DZ} find respectively
\begin{equation}
\ell(\ell+1)C_\ell\left|_{{~}_{\!\! SW}} \sim 2 \epsilon^2
\right. ~~\mbox{  and } ~~
\ell(\ell+1)C_\ell\left|_{{~}_{\!\! SW}} \sim 8 \epsilon^2.
\right. \label{SW}
\end{equation}
According to  Refs.~\cite{BR,PST}, the Doppler peak is a factor of $\sim 3.4$
times higher than the SW plateau, whereas it is only about 1.5 times higher if
the result found in \cite{DZ} is assumed. (We allow for about 30\% of the SW
amplitude to be added in phase to the Doppler amplitude of
$\sim 2.24\epsilon$,
according to Eq.~(\ref{clfirstpeak})). Clearly, improved
numerical simulations
or analytical approximations are needed to resolve this discrepancy.
However, it is apparent from Eqs.~(\ref{clfirstpeak}) and
(\ref{SW})
that the Doppler contribution from textures is somewhat
 smaller than for generic inflationary models.

We believe that our results are basically valid for all global
defects. This depends crucially on the $ 1/\sqrt{t}$ behavior of
$(\phi')^2$ on large scales (cf. Eq.~(\ref{fit})),
which is a generic feature of global defects:
on super--horizon scales,
$(\phi')^2(k)$ represents white noise superimposed on the average given by
 $(\phi')^2(k=0) \propto \sqrt{V}/t^2$. Since there are
$N=V/t^3$ independent patches in a simulation volume $V$,
the amplitude of $(\phi')^2(k)$
is proportional to $\sqrt{V}/(t^2\sqrt{N}) \propto 1/\sqrt{t}$.
(Notice that this argument does not apply for local cosmic strings.)

Based on our analysis, we conclude that
if the existence of Doppler peaks is indeed confirmed
and if the first peak is positioned at $\ell<300$, {\sl then}
global topological defects are ruled out.
On the other hand, if the first Doppler peak is positioned at
$\ell \sim 350$ and if its amplitude
is lower than  the one predicted for standard inflationary models, global
topological defects
are strongly favored if compared to the latter.
(There are however non--generic, open, tilted inflationary models which
might reproduce a similar signature in the CMB angular power spectrum).
To our knowledge this is the first clear fingerprint within present
observational capabilities, to distinguish among these
two competing models of structure formation.

As a future work, we aim to model with better accuracy the global scalar
field $\phi$ during both the radiation and the matter dominated era, as well
as to include the SW effect and the photon diffusion.  This will allow us
to better estimate the amplitude of the first Doppler peak and to investigate
secondary peaks.

As we were completing our work, a preprint\cite{ct}
on the same issue, but following a different approach,  came to our
attention.
The authors calculate the Doppler peaks from cosmic textures in
the synchronous gauge. They include the Sachs Wolfe contribution
into the analysis, but they need more of the uncertain modeling of the defect
stress energy tensor. Even though we  basically agree
with the shape and position of their Doppler peaks, we obtain a somewhat
smaller amplitude.
\vspace{1cm}\\
{\bf Acknowledgement}\\
We thank Leandros Perivolaropoulos, who participated in the
beginning of this project and Mark Hindmarsh, for helpful discussions and in
particular for his skills with MATLAB. One of us (R.D.) acknowledges
stimulating discussions with Neil Turok. A.G. thanks Nuno Antunes and
Dennis Sciama for
encouragement, the Institut f\"ur Theoretische Physik, Z\"urich for hospitality
and The British Council for partial financial support.
This work was partially supported by the Swiss NSF.

\end{document}